\documentclass{aa}  
\usepackage{graphicx}
\usepackage{txfonts}
\begin{document}
   \title{Modeling the UBVRI time delays in Mrk 335}

   \author{B. Czerny
          \inst{1}
          \and
          A. Janiuk\inst{2}
          }

   \offprints{A. Janiuk}

   \institute{Copernicus Astronomical Center, Bartycka 18, Warsaw, Poland\\
         \and
        Department of Physics, University of Nevada, Las Vegas, NV89154, USA\\
             \email{ajaniuk@physics.unlv.edu, bcz@camk.edu.pl}
             }

   \date{Received ??? Accepted ???}

 
  \abstract
   { }
   {We develop a model of time delays between the continuum bands in the Narrow 
   Line Seyfert 1 galaxy Mrk 335 to explain the observed delays measured in this source.}
   {We consider two geometries: an accretion disk with fully ionized warm absorber of considerable optical depth, located close to the symmetry axis, and an accretion disk with a hot corona. Both media lead to significant disk irradiation but the disk/corona geometry gives lower values of the time delays. }
   {Only the disk/corona models give results consistent with measurements of Sergeev et al., and a low value of the disk inclination is favored. The presence of an optically thick, fully ionized outflow is ruled out at the 
2-$\sigma$ level. }
  {}


  \authorrunning{B. Czerny and A. Janiuk}
  \titlerunning{Modeling the UBVRI time delays in Mrk 335}
   \maketitle
%

\section{Introduction}

Active galactic nuclei (AGN) are powered by accretion onto a central massive 
black hole, and the accretion is believed to proceed predominantly through
the accretion disk. However, the shapes of the broad band spectra of most AGN
are much broader than  predicted by the classical disk model of Shakura
\& Sunyaev (1973). The deviation is not as strong in the case of bright 
quasars (e.g. Zheng et al. 1997, Laor et al. 1997, Koratkar \& Blaes 1999, 
Czerny et al. 2004, 
Shang et al. 2005), but in the case of Seyfert 1 galaxies
the role of the X-ray emission is essential and the presence of the accretion
disk is far less obvious when looking at the optical data. The spectral slopes
are much redder than the standard slope of $p = 1/3$ of the Shakura \& Sunyaev (1973)
disk in the
$F_{\nu} \propto \nu^{p}$ representation (e.g. 
Edelson \& Malkan 1986; Kong et al. 2004). The analysis of
the X-ray spectral features may support the view
that the disk is disrupted in the inner part of the flow 
(a relatively narrow iron K$\alpha$ is line preferred, e.g. Yaqoob et al. 2001, 
Reeves et al. 2004, although a weak broad component cannot be ruled out 
with high significance, e.g. Yaqoob \& Padmanabhan 2004, Yaqoob et al. 2005).

Narrow Line Seyfert 1 galaxies (NLS1) are an intermediate type of objects: 
their spectra are
dominated by disk emission, as in bright quasars (see e.g. PG1211+143,
Czerny \& Elvis 1987; RE J1034+396, Puchnarewicz et al. 2001; Mrk 335, Pounds et 
al. 1987), and the disk is likely to extend down to the marginally stable orbit 
(as inferred from the iron line shape in MCG -6-30-15, Tanaka et al. 1995, Fabian et al. 
2002; 1H0707-495, Fabian et al. 2004; PG1211+143, Janiuk et al. 2001, 
IRAS 13349+2438, Longinotti et al. 2003) but the 
deviations from the standard disk
are quite strong as well. NLS1 galaxies usually have profound soft X-ray excesses 
and/or steep (soft) X-ray spectra (e.g. Brandt et al. 1997, Leighly 1999, Romano 
et al. 2004, Boller 2004), and their X-ray spectra frequently 
show warm absorber features (e.g. in MCG-6-30-15, Turner et al. 2004; 
NGC 4051, Collinge et al. 2001, Ton S 180, 
R\' o\. za\' nska et al. 2004; PG 1211+143, Pounds et al. 2003). It is therefore interesting to explore which 
additional component, apart from an accretion disk, might be responsible for 
the overall spectral shape. 

We select one of the most studied NLS1 galaxies, Mrk 335 ($z = 0.026$),
for the modeling. This is a typical NLS1 object, with a narrow 
$H\beta$ line (1640 km s$^{-1}$, Wang et al. 1996), and variable 
Big Blue Bump 
with a large soft X-ray excess (Edelson \& Malkan 1986, Pounds et al. 1987, 
Turner \& Pounds 1988, Turner et al. 1993, Nandra \& Pounds 1994, 
Leighly 1999), possibly a 
relativistically smeared iron K$\alpha$ line 
(Gondoin et al. 2002, Crummy et al. 2006), and direct evidence of a
weak partially ionized warm absorber (Turner et al. 1993, Reynolds 1997, 
Bianchi et al. 2001). Since the optical/UV slope is roughly a power
law ($F_{\nu} \propto \nu^{\alpha}$, with $\alpha = -0.64 \pm 0.01$, Constantin
\& Shields 2003) and does not constrain the model parameters very strongly, 
we concentrate
here on modeling the observed time delays between the UBVRI bands measured
for this object. The source was monitored for several years at the Crimean 
observatory and the
time delays between the UBVRI bands were measured (Doroshenko et al. 2005,
Sergeev et al. 2005). We describe the model in Sect.~\ref{sect:model}.
In Sect.~\ref{sect:results} we describe its application to Mrk~335, and
in Sect.~\ref{sect:discussion}
we discuss the consistency of the results with independent estimates of the
central black hole mass and the optical depth of the warm absorber. 

\section{Model}
\label{sect:model}

The UV/optical/IR continuum of Mrk 335 varies significantly (Sun et al. 1994,
Kassebaum et al. 1997, 
Peterson et al. 1998, Tao et al. 2004, Klimek et al. 2004). The measured delays
of the continuum between the U band and BVRI bands increase with the 
wavelength difference (Sergeev et al. 2005, Doroshenko et al. 2005).
Such variations in AGN spectra are most likely caused by the irradiation of the outer parts of the accretion
disk responsible for the optical/UV emission by the EUV and X-ray radiation 
generated in the innermost part of the disk, as discussed by
Collin-Souffrin (1991) (see also Courvoisier \& Clavel 1991, Krolik et al. 
1991, Collier et al. 1998 and Oknyanskij et al. 2003).
The direct irradiation of the outer parts of the disk
by the inner disk is unlikely (Loska
et al. 2004). The reason is that AGN disks are geometrically very thin, much
thinner than GBH disks. In Seyfert 1 galaxies the irradiation is possible due to 
the extended optically thin X-ray emitting inner region, which increases the
geometrical factor (Rokaki et al. 1993, Chiang 2002). However, in NLS1 the
standard, optically thick disk extends to the marginally stable orbit,
 rather than being disrupted in its inner part, and therefore the
presence of an optically thin inner flow is rather
unlikely. 

Therefore, we suggest here that additional
material may efficiently redirect the nuclear emission towards the outer
disk. We consider in detail two geometries: a warm absorber type medium, 
predominantly 
occupying the region around the symmetry axis,
 and the disk wind/corona, located
closer to the disk plane. 

   \begin{figure}
   \includegraphics[width=8.5cm]{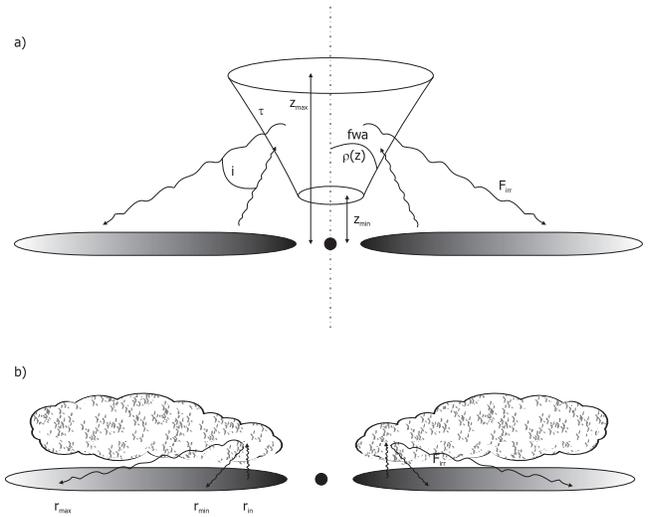}
      \caption{The two geometries used: a disk/warm absorber geometry (panel a) and disk/corona geometry (panel b). Differences in the location of the scattering medium lead to differences in the time delays between the continuum bands.  
              }
         \label{fig:geom}
   \end{figure}

\subsection{Disk with a warm absorber}

We apply here the simple model of disk irradiation described in Czerny
et al. (2003). We neglect the true absorption effect due to the warm absorber
since (i) we are mostly interested in the fully ionized fraction of this 
material (ii) the absorption is limited to the 0.5 - 2 keV energy band, and the 
scattering effect (working in the whole energy band) is more effective than 
absorption and reemission in redirecting outgoing photons towards the disk. 
We assume that most of the scattering medium is
located near the symmetry axis in the form of a cone (see Fig.~\ref{fig:geom}). 
The warm absorber 
parameters are: the fraction of the sky covered by the cone, $f_{wa}$ 
(equivalently, the
opening angle of the cone), the smallest and the largest distance, $z_{min}$
and $z_{max}$, 
of the scattering material measured
along the symmetry axis, the parameter $\delta$ that measures the density
distribution of the warm absorber ( $\rho(z) \propto z^{-\delta}$), and
the total optical depth of the medium for scattering, $\tau$.

\begin{figure}
   \includegraphics[width=8.5cm]{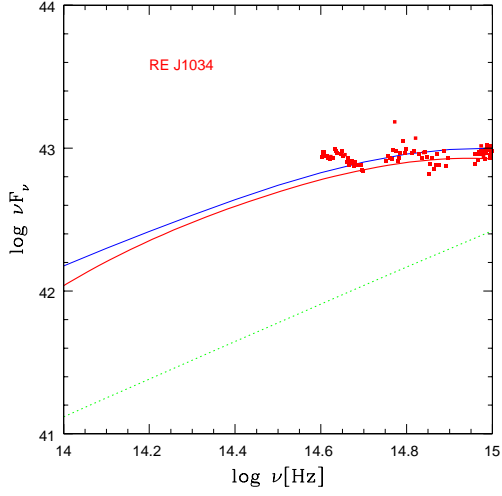}
      \caption{Comparison of the full 3-D irradiation computations with self-irradiation included (Loska et al. 2004, thin continuous line) with the results of the simplified scheme used here (thick continuous line) in the case of extreme  
illumination required by the source RE J1034. The bare disk model is marked with a dotted line. Parameters: $M = 6.3 \times 10^5 M_{\odot}$, $\dot m = 0.84$, 
$\delta = 0.0$, $\tau = 0.6$, $z_{min}= 10 R_{Schw}$, $z_{max}=7000 R_{Schw}$, 
$f_{wa} = 0.5$. 
              }
         \label{fig:j1034}
   \end{figure}

The local irradiation flux is given by
\begin{equation}
F(r) =  {\int_{z_{min}}^{z_{max}} {\cal F}(r,z) dz},
\end{equation}
where 
\begin{equation}
{\cal F}(r,z) = {\tau f_{wa} (\delta-1) L \over z_{min}^{1-\delta} - z_{max}^{1-\delta}} {z^{1-\delta} \over (z^2 + r^2)^{3/2}},
\end{equation}
with $L$ being the total luminosity of the source.
This approach is much less time consuming than the more accurate modeling
done by Loska et al. (2004). 
However, our model is still capable of producing reliable results. In the most
extreme case of irradiation considered for the source RE J1034+396, the 
simplified numerical scheme underestimates the flux in the optical range by
25 \% (see Fig.~\ref{fig:j1034}) but the overall shape is preserved; in 
moderate illumination cases, with higher
$\delta$, the error does not exceed a few percent.
 
The disk internal emission is described in the Newtonian approximation, and
parameterized by the black hole mass, $M$, and the dimensionless accretion 
rate, $\dot m$, defined with the efficiency factor of 1/12 included (appropriate for Newtonian boundary condition, Shakura \& Sunyaev 1973), i.e.
\begin{equation}
\dot m = {\dot M \over \dot M_{Edd}}; ~~~~~~ \dot M_{Edd}={48 \pi GM m_p\over c \sigma_T}. 
\end{equation} 

The illuminated disk is assumed to radiate locally as a black body. 
This simplifying assumption allowed us to save computer time, 
and is fully justified. However more detailed models of the disk emission do show 
a departure 
of the local spectrum from a black body, but this is only if the opacities 
of heavy 
elements are neglected.
As shown e.g. in Hubeny et al. (2001) and
Madej \& R\' o\. za\' nska (2000), taking these opacity effects 
into account minimizes the difference with respect to a blackbody spectrum
(apart from
the soft X-ray part).

The emitted disk spectrum passes through the warm absorber, so the observed
spectrum is reduced by the factor $\exp(-\tau)$ with respect to the intrinsic 
spectrum. In this way we neglect the possible clumpiness of the warm absorber,
which in principle may decouple the specific line of sight from the average
scattering probability. 

The time delays are calculated assuming that, for a given radius, all 
irradiating flux comes from
a single representative distance on the symmetry axis. This distance is 
calculated as the effective distance, $z_{eff}$ in the following way
\begin{equation}
z_{eff} = {\int_{z_{min}}^{z_{max}} {\cal F}(r,z) z dz \over 
\int_{z_{min}}^{z_{max}} {\cal F} (r,z) dz}.
\end{equation}
The delay of the emission from a given radius with respect to the emission
from the innermost part of the disk is calculated as
\begin{equation}
\Delta t (r) = r \sin i + z_{eff} + (z_{eff}^2 + r^2)^{1/2}.
\end{equation}

\subsection{Disk with a flaring corona}

\begin{figure}
   \includegraphics[width=8.5cm]{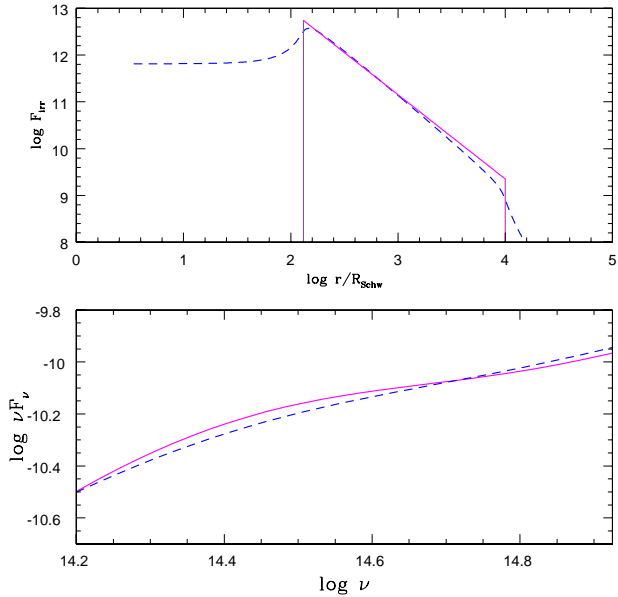}
      \caption{Comparison of the 3-D irradiation computations of scattering by the corona with radial density $\rho \propto r^{-0.8}$, extending from $r_{in}=130 R_{Schw}$ to $r_{out}= 10^4 R_{Schw}$, total optical depth in the radial direction 0.734 and constant opening angle of $15^{\circ}$ measured from the equatorial plane (dashed line) with the results of the simplified scheme used in the present paper (continuous line) for equivalent parameters, i.e. the same $r_{in}$, $r_{out}$, black hole mass and accretion rate, $M = 10^7 M_{\odot}$, $\dot m = 0.95$, but with $\gamma = 1.8$ and $f = 0.095$. Such a value of the factor $f$  is equivalent to the scattering probability given by the optical depth and the azimuthal extension of the corona in numerical computations. It can be shown analytically that the power law distribution of the coronal density with a given slope leads to a disk region irradiated by the flux decreasing as a power law with index larger by 1. 
              }
         \label{fig:test2}
   \end{figure}

The accretion disk is also likely to develop an extended corona or a 
wind-type outflow. In this case the additional hot material is located
predominantly close to the disk surface but high enough to intercept
some of the nuclear flux much more effectively than the disk surface itself. There
may be some dissipation taking place directly in the wind-corona. Both
effects will increase the local disk temperature. Since there are no
well justified models of this situation, we model such a disk
assuming that a fraction of the disk, between $r_{min}$ and $r_{max}$,
is irradiated by additional flux. This irradiating flux is simply
parameterized as a power law
\begin{equation}
F_{irr} = A \left({r \over r_{min}}\right)^{\gamma},
\end{equation}
where the normalization coefficient A is introduced through the parameter $f$
describing the ratio of the total energy in this component to the total
energy due to accretion
\begin{equation}
A = {f L \over 4 \pi r_{min}^2\left[{1 \over 2 - \gamma}\left({r_{max} \over r_{min}}\right)^{2 - \gamma} - 1\right]}.
\end{equation}
Again, we tested the analytical approach against the 3-D numerical computations
of the scattering in a corona which has a density profile in the radial 
direction but is uniform in the $\theta$ direction (in spherical coordinates) and has a
constant opening angle. The results are illustrated in Fig.~\ref{fig:test2}. Our analytical model is approximately correct although the exact results predict some irradiation effect in the innermost part of the disk (below $r_{min}$, roughly constant with radius) as well as above $r_{max}$ (decreasing roughly as $r^{-3}$). The irradiation is significantly different from a point-like source since in our model we have an extended region with incident flux decreasing as an arbitrary power of radius (physically determined by the density distribution in the corona) while in the case of a point-like irradiation the incident flux is always constant below a certain radius and decreases as $r^{-3}$ above it.

In computations of time delays we neglect the extension of the coronal medium 
in the vertical direction, so the emission from a given radius is delayed with respect to the central
regions by
\begin{equation}
\Delta t (r) = r (\sin i + 1).
\end{equation}

\subsection{Time delays in UBVRI bands}

In order to determine the time delay at a given wavelength with respect to the
emission from the innermost part of the disk we have to connect a given 
radius with a 
representative wavelength. In the simplest approximation, this can be done
taking into account that a local black body emission with the temperature $T$ 
peaks at $2.7 kT$ in a $\log F_{\nu}$ vs. $\log {\nu}$ diagram. However, if the
spectrum is not of a single black body type, the dominant contribution 
may not be well
determined by such a criterion (for example, for a power law spectrum a factor 
before  $kT$ depends on the spectral slope; see Siemiginowska \&
Czerny 1989). Since the shapes of the model spectra are more complex than a 
single power law, we use a numerical approach. In order to connect a given 
radius with a representative wavelength, we calculate a model with a narrow 
gap in the disk around a certain radius, then we compare the spectrum to the 
model without a gap. We find the wavelength where the ratio of the 
gap model spectrum
to the full model spectrum was the lowest. By repeating such a procedure we 
identify a set of radii with a corresponding sequence of wavelengths.

\subsection{Data}

The Galactic hydrogen column in the direction of Mrk~335 is 
$N_H = 3.8 \times 10^{20}$ cm$^{-2}$ (Veron-Cetty et al. 2001). The reddening 
can therefore be estimated by taking the standard relation for the interstellar
medium in the Galaxy
$N_H/E(B-V)=5.8 \times 10^{21}$ (Bohlin et al. 1978) which gives $E(B-V) = 
0.066$ and the data were corrected accordingly.  The continuum fluxes used for
model fitting are given in Table~\ref{tab:continuum}.

\begin{table}
\caption{Continuum data points of Mrk 335 used for model fitting}
\begin{center}
\begin{tabular}{l r }     
\hline\hline     
 $\log \nu$ & flux [mJy] \\ 
\hline 
U    & 14.23        \\    
B    & 13.99        \\    
V    & 15.51        \\
R    & 18.09         \\
I    & 19.49	     \\
J1   & 21.00         \\
J2   & 22.10         \\
\hline                  
\end{tabular}
\end{center}
The data points are taken from NED (McAlary et al. 1983, Spinoglio et al. 
1995). 
\label{tab:continuum}
\end{table}

The source was systematically monitored in Crimea for 10 years (Doroshenko et 
al. 2005, Sergeev et al. 2005). The measured delays between the various optical bands are given in Table~\ref{tab:delays}.

\begin{table*}
\caption{Measured time delays in Mrk 335}
\begin{center}
\begin{tabular}{l r r r r}     
\hline\hline     
 Band & Doroshenko et al.& Doroshenko et al.& Sergeev et al. &  Sergeev et al.\\ 
      &   peak      & centroid     &   peak            &   centroid      \\ 
\hline
U    & -    & -     \\
B    & $0.4^{+3.3}_{-0.8}$ & $2.2^{+6.8}_{-3.4}$  & -  & -   \\
V    & $6.0^{+8.8}_{-0.9}$ & $4.5^{+14.5}_{-1.0}$  & $0.43^{+0.83}_{-0.81}$ &  $0.86^{+0.56}_{-0.97}$    \\
Rc   & $5.0^{+12.1}_{-2.1}$ & $8.3^{+15.1}_{-2.5}$     & $2.30^{+0.44}_{-1.76}$ &  $1.45^{+0.94}_{-1.26}$    \\
I/Ic & $7.6^{+11.2}_{-1.9}$ & $7.6^{+15.7}_{-2.9}$    & $2.33^{+0.30}_{-1.86}$ & 	$1.74^{+0.78}_{-1.25}$  \\
\hline                  
\end{tabular}
\end{center}
The delays measured with respect to the U band in Doroshenko et al. (2005) and with respect to the B band in Sergeev et al. (2005). Either the standard or Cousins filter system was used (Rc,Ic). Quoted measurement uncertainties are 1 $\sigma$ errors. 
\label{tab:delays}
\end{table*}

The measured delays generally depend on the method (peak or centroid), 
and the errors in measurements with respect to the U band are much larger than 
those measured with respect to the B band. The I vs. B delay of Sergeev et al. 
(2005) gives an upper 1 $\sigma$ limit of 2.5 - 2.6 day, independent of the 
method.
 
The delays measured by Doroshenko et al. (2005) are systematically larger. All
models are consistent with 1 $\sigma$ upper limits of more than 17 days of 
delay between the U and I band. However, we consider this delay measurements 
as much less certain than the delays given by Sergeev et al. (2005) since the
measurement of U flux is much less certain than the measurement of B, when a 
relatively small telescope located in a poor astronomical site is used.




\section{Results}
\label{sect:results}

   \begin{figure}
   \includegraphics[width=8.5cm]{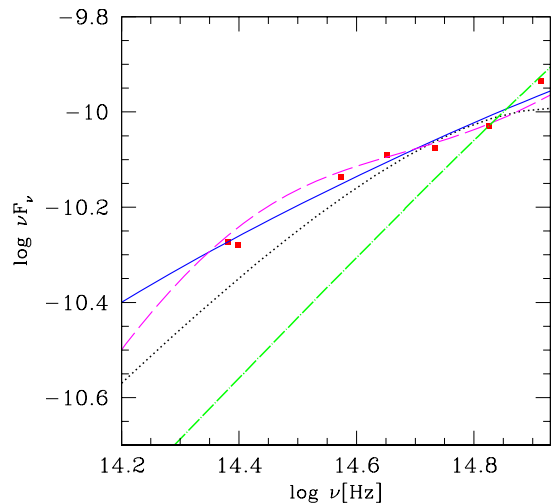}
      \caption{Representation of the optical data by two bare disk 
       models: $M = 7 \times 10^8 M_{\odot}$, $\dot m = 0.0015$ 
      (dotted line), and $M = 5 \times 10^7 M_{\odot}$, $\dot m = 0.1$ 
      (long dash-dot line), an example of the fit with a warm absorber: 
      $M = 1.78 \times 10^7 M_{\odot}$, $\dot m = 0.6$, $\delta = 1.4$, 
      $\tau = 0.8$, $z_{min}= 50 R_{Schw}$, $z_{max}=40000 R_{Schw}$, 
      $\cos i = 1$  (continuous line), and an example of the fit with corona: 
      $M = 1.0 \times 10^7 M_{\odot}$, $\dot m = 0.95$, $\gamma = 1.8$, 
     $f = 0.095$, $r_{min}= 130 R_{Schw}$, $r_{max}=10000 R_{Schw}$, 
     $\cos i = 1$ (long dashed line).
The solid squares mark the data points (cf. Table \ref{tab:continuum}). 
              }
         \label{fig:no_wa}
   \end{figure}

\subsection{Bare disk model}

The presence of time delays in Mrk 335 between the optical bands, increasing 
with the wavelength separation, clearly suggests the role of reprocessing in 
the formation of the optical spectra. However, as a reference we consider a 
simple fit to the
spectrum by a non-irradiated standard disk. 
	
The spectrum is relatively blue at longer wavelengths ($p \approx -0.2$), but
it is still not as steep as expected from the standard disk, and it flattens
at shorter wavelengths to $p \approx -0.5$. Such a curved spectrum can be
fitted by a standard disk if the black hole mass is high, so the maximum
disk temperature is within the optical/UV band. 
An example of the data fit without the presence of the warm absorber is shown
in Fig.~\ref{fig:no_wa} (dotted line). The model does not fit the near-IR 
data points but
this can be explained by starlight contamination. However, such a model 
cannot account for the total 
luminosity of the source (the model bolometric luminosity is only  
$1.3 \times 10^{44}$ erg s$^{-1}$ while the source bolometric luminosity was 
estimated to be much higher, $1.3 \times 10^{45}$ erg s$^{-1}$; Pounds 
et al. 1987, or $6 \times 10^{45}$ erg s$^{-1}$; Edelson \& Malkan 1986). The 
required mass is much higher than the value obtained from the reverberation
($M = 1.42 \pm 0.37 \times 10^7 M_{\odot}$, Peterson et al. 2004). 

On the other hand, if we fix the accretion rate at some larger value, as
expected for NLS1 galaxies, and we adjust the black hole mass to the required
emission level, the resulting black hole mass is not so large (e.g. $M = 5 
\times 10^7 M_{\odot}$ for adopted $\dot m = 0.1$) but the predicted spectrum
is much too blue to represent the data points (see the steepest line in 
Fig.~\ref{fig:no_wa}).

Therefore, the bare disk model
is unlikely to represent well the Mrk 335 data.  Most of the older models
based on the fit to optical/UV data alone would now face the same problem,
as the masses inferred by such models are high (e.g. Sun \& Malkan 1989, $M = 8.9 \times 
10^7 M_{\odot}$ for a top view, and it is larger for higher inclination angles; 
Zheng et al. 1995, $M = 5 \times 10^7 M_{\odot}$). 

\subsection{Irradiated disk model}

Both geometries, irradiation due to the scattering by a fully ionized warm 
absorber and due to scattering by the disk corona, can account for the 
overall shape of the spectrum. We did not try to differentiate between the 
models on the basis of the quality of the fit because the observational points
can have some systematic errors due to slight internal reddening, starlight
contamination and the contribution of the small blue bump (i.e. Fe II and Balmer continuum) to the continuum.
Also the theoretical predictions are not precise; various treatments of
the radiative transfer in the disk surface can lead to differences between
the models of the order of 20\% in the U band if the normalization in the I band is fixed 
(see e.g. Ross et al. 1992, Hubeny et al. 2001).
Since for each mass, accretion rate and inclination angle we fit other model
parameters, we obtain a satisfactory data representation for a broad range of 
these basic parameters. The results for the time delays 
for acceptable values of mass and accretion rates 
are given in Table~\ref{tab:cosi1} and 
Table~\ref{tab:cosi2} for the two models: the warm absorber and disk/corona 
system, respectively.
  
In both cases the acceptable solutions create an extended elongated region in
 $M - \dot m$
parameter space since an increase in the black hole mass can be compensated 
by the decrease in the accretion rate in order to normalise 
the spectrum consistently with the data (see for example the discussion by 
Tripp et al. 1994 of the bare disk data fitting). The location of this 
region depends 
on the inclination angle: larger disk inclinations allow for larger masses 
for a given
accretion rate. The width of this elongated region is provided by the 
additional
model parameters,  and it is much larger in the disk/warm absorber scenario than in 
the disk/corona scenario. For example, if we fix the black mass
at $10^7 M_{\odot}$ and the inclination angle at $\cos i = 1.0$, 
the acceptable fits in the disk/warm absorber geometry are
obtained for $\dot m = 0.6$ ($\delta = 1.4$, $\tau = 0.25$,
$f_{wa} = 1.0$, $z_{min}= 130 R_{Schw}$, $z_{max}=40000 R_{Schw}$), for
$\dot m = 0.8$ ($\delta = 1.4$, $\tau = 0.30$,
$f_{wa} = 0.6$, $z_{min}= 130 R_{Schw}$, $z_{max}=10000 R_{Schw}$), and
for $\dot m = 0.8$ ($\delta = 1.6$, $\tau = 0.15$,
$f_{wa} = 0.6$, $z_{min}= 370 R_{Schw}$, $z_{max}=40000 R_{Schw}$). In the disk/corona geometry, the acceptable fits are for $\dot m = 0.8$ 
($\gamma = 1.8$, 
$f = 0.13$, $r_{min}= 50 R_{Schw}$, $r_{max}=10000 R_{Schw}$) and for
$\dot m = 1.0$ 
($\gamma = 1.8$, 
$f = 0.095$, $r_{min}= 130 R_{Schw}$, $r_{max}=10000 R_{Schw}$).

We give in Table~\ref{tab:cosi1} and 
Table~\ref{tab:cosi2} the theoretical delays between the B and I band. The 
delays are systematically larger for the disk/warm absorber geometry than 
for the disk/corona geometry since the light travel time through the warm 
absorber and back to the disk contributes to the delays.

\begin{table}
\caption{Time delays between B and I band in days from the disk/warm absorber 
system}
\begin{center}
\begin{tabular}{l r r r r r r r }     
\hline\hline     
log M:  & 6.5  & 6.75  &  7.0 &  7.25 &  7.5 &  7.75 & 8.0 \\  
\hline
\multicolumn{8}{c}{cos i = 1.0}\\
\hline
log $\dot m$ \\ 
\hline
0.2    & x    & x     &  x    &  x    &  3.0 &  4.3  & x   \\
0.4    & x    & x     &  x    &  3.5  &  3.6 &  7.7  & x   \\
0.6    & x    & x     &  3.7  &  6.1  &  5.6 &  x    & x   \\
0.8    & x    & x     &  3.2  &  3.3  &  6.2 &  8.3  & x   \\
1.0    & x    & 2.9   &  3.0  &  4.5  &  x   &  x    & x   \\
\hline                  
\multicolumn{8}{c}{cos i = 0.75}\\
\hline
log $\dot m$ \\ 
\hline
0.2    & x    & x     &  x    &  x    &  6.0 &  4.4  & 9.3   \\
0.4    & x    & x     &  x    &  5.2  &  9.7 &  8.0  & 11.0  \\
0.6    & x    & x     &  5.1  &  4.6  &  8.7 &  x    & x     \\
0.8    & x    & x     &  6.1  &  9.3  &  7.8 &  x    & x     \\
1.0    & x    & x     &  7.0  &  6.3  &  9.5 &  x    & x     \\
\hline                  
\multicolumn{8}{c}{cos i = 0.50}\\
\hline
log $\dot m$ \\ 
\hline
0.2    & x    & x     &  x    &  x    &  x   &  8.2  & 12.0   \\
0.4    & x    & x     &  x    &  x    &  7.3 &  9.0  & 13.0  \\
0.6    & x    & x     &  x    &  6.6  & 10.0 & 12.0  & 14.0   \\
0.8    & x    & x     &  7.6  &  6.9  & 13.0 & 14.0  & 17.0  \\
1.0    & x    & x     &  5.9  &  6.7  & 17.0 & x     & x     \\
\hline                  
\end{tabular}
\end{center}
Accretion rate is given in dimensionless units, assuming 
standard efficiency of
accretion onto a non-rotating black hole. Other model parameters
were found as the best fit to the optical/IR data of Mrk 335. Signs 'x' mark the values of 
black hole masses
and accretion rates for which no acceptable model can be fitted to
the data. 
\label{tab:cosi1}
\end{table}

\begin{table}
\caption{Time delays between B and I band in days from the disk/corona 
system}
\begin{center}
\begin{tabular}{l r r r r r r r }     
\hline\hline     
log M:  & 6.5  & 6.75  &  7.0 &  7.25 &  7.5 &  7.75 & 8.0 \\  
\hline
\multicolumn{8}{c}{cos i = 1.0}\\
\hline
log $\dot m$ \\ 
\hline
0.2    & x    & x     &  x    &  x    &  1.7 &  x    & x   \\
0.4    & x    & x     &  x    &  2.1  &  x   &  x    & x   \\
0.6    & x    & x     &  x    &  1.9  &  x   &  x    & x   \\
0.8    & x    & x     &  1.9  &  x    &  x   &  x    & x   \\
1.0    & x    & x     &  2.0  &  x    &  x   &  x    & x   \\
\hline                  
\multicolumn{8}{c}{cos i = 0.75}\\
\hline
log $\dot m$ \\ 
\hline
0.2    & x    & x     &  x    &  x    &  x   &  x    & x   \\
0.4    & x    & x     &  x    &  x    &  3.3 &  x    & x  \\
0.6    & x    & x     &  x    &  3.7  &  x   &  x    & x     \\
0.8    & x    & x     &  x    &  3.7  &  x   &  x    & x     \\
1.0    & x    & x     &  3.7  &  3.7  &  x   &  x    & x     \\
\hline                  
\multicolumn{8}{c}{cos i = 0.50}\\
\hline
log $\dot m$ \\ 
\hline
0.2    & x    & x     &  x    &  x    &  x   &  4.0  & x   \\
0.4    & x    & x     &  x    &  x    &  x   &  x    & x  \\
0.6    & x    & x     &  x    &  x    & 5.3  &  x    & x   \\
0.8    & x    & x     &  x    &  5.8  & x    &  x    & x  \\
1.0    & x    & x     &  4.5  &  5.3  & x    &  x    & x     \\
\hline                  
\end{tabular}
\end{center}
Accretion rate is given in dimensionless units, assuming 
standard efficiency of
accretion onto a non-rotating black hole. Other model parameters
were found as the best fit to the optical/IR data of Mkr 335. Signs 'x' mark the values of 
black hole masses
and accretion rates for which no acceptable model can be fitted to
the data. 
\label{tab:cosi2}
\end{table}

We compare these delays to the observations summarized
 in Table~\ref{tab:delays}.
Only predictions of the disk/corona geometry at low inclination ($\cos i = 1$) 
are consistent with the {upper 1 $\sigma$ limit of Sergeev et al (2005)}
These models cover a broad range of accretion rates (from 0.2 to 1 in 
dimensionless units)
and black hole masses range from $10^7 M_{\odot}$ to $3 \times 10^7 M_{\odot}$.

\begin{figure}
   \includegraphics[width=8.5cm]{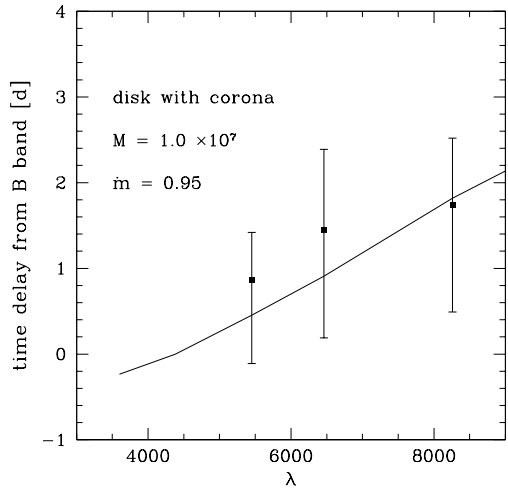}
      \caption{Time delays measured with respect to the B band. Data points: centroid delays from Sergeev et al. (2005). Model: disk/corona geometry with parameters as in Fig.~\ref{fig:no_wa}.
              }
         \label{fig:delays}
   \end{figure}

An example of the wavelength-dependent time delays for one of the 
disk/corona models is shown in Fig.~\ref{fig:delays}. We see a good 
agreement between the model and the data. The delays increase with
the wavelength separation in a similar way in the model and in the data. 

None of the disk/warm absorber models is acceptable within a 1 $\sigma$ error. 

If a larger 3 $\sigma$ error is allowed, the upper limit for the I vs. B delay 
increases to 7.1 days (centroid measurement) and to 7.0 day (peak measurement).
 All disk/corona models, as well as several low inclination examples of 
the disk/warm absorber geometry, satisfy this requirement. 

Although the numerical method allows us to determine the time delays precisely
from the point of view of the adopted algorithm, the actual errors of such 
determinations are significant. Using the method outlined above, we tested
the radial extension of the region responsible for emission at a given 
wavelength. For a specific model, the radial extension of the zone responsible
for 50\% of the flux in B band was from $250 R_{Schw}$ to $1260 R_Schw$,
and the zone responsible for  50\% of the flux in I band lay between 
$560R_{Schw}$ and $4400 R_{Schw}$. It is
therefore not surprising that the errors of the measured time delays are also
large.  

We also checked whether the dependence of the variability level on the 
wavelength is well reproduced by any of the models. We considered two luminosity states in a single representative model: one with the accretion rate (and luminosity) enhanced in the innermost 10 $R_{Schw}$ and one with a suppressed innermost accretion rate. The enhancement/suppression factor was adjusted to the ratio of the minimum to maximum flux in B band given by Doroshenko et al. (2005). We used their photometric measurements for that purpose. The basic model parameters were as in Fig.~\ref{fig:no_wa}.

We give the results in Table~\ref{tab:rms} for the disk/corona model, but the 
results for the disk/warm absorber geometry are almost the same. In the data, 
the variability amplitude decreases towards IR but in the model it actually 
increases. This is due to the fact that we assumed a constant intrinsic disk 
emission at larger radii and variable irradiation, and that the role of irradiation 
increases towards the IR in order to reproduce the observed spectra that are much redder 
than the intrinsic disk spectrum. We can assume that a certain amount of 
starlight contributes to the spectrum in the VRI band. However, the amount of 
starlight required to suppress the variability is very large (see column 4 in 
Table~\ref{tab:rms}). If the spectrum is corrected for such an amount of 
starlight the intrinsic spectrum becomes bluer than in the standard disk, leaving 
no space for the effect of irradiation and time delays. We cannot perform 
self-consistent computations for the full model, with starlight iterations 
(i.e adopting a certain starlight distribution, finding new best solution for 
the spectrum, checking the variability properties, correcting the starlight 
again etc), 
since this is too time consuming. However, we estimated the probability of
finding a consistent solution satisfying both the requirement of a red spectral
slope and variability amplitude decreasing strongly with increasing
wavelength. We used for that purpose an analytical power law description for
all three spectral component independently, assuming an arbitrary power law
index for the starlight contribution, an index of 1/3 for the disk, an index
of -1 for the spectral component due to irradiation, and we searched the
parameter space of the starlight index and the two normalization factors, 
determining starlight to disk and irradiation to disk ratios. In all solutions
with variability larger in B band than in I band, the difference in the 
variability level was by 1 - 2 \%, the IR slope in the final spectrum was 
unacceptably steep (with power law index above 2.7), and the overall 
variability level in B band was too low ($F_{max}/F_{min} \sim 1.1$) for models
with the IR index smaller than 3.0.

The 
variability at longer wavelengths may be smeared, and the amplitude effectively 
lowered due to the larger reprocessing area. More advanced time-resolved 
modeling is necessary to verify this possibility.  

\begin{table}
\caption{The variability properties in the BVRI band}
\begin{center}
\begin{tabular}{l r r r}     
\hline\hline     
 & $F_{max}/F_{min}$ & $F_{max}/F_{min}$ & starlight \\
 & measured          & disk/corona model  & required \\
\hline 
B    & 1.550    &   1.550   &  0  \\    
V    & 1.384    &   1.667   &  36  \\
R    & 1.358    &   1.703   &  42   \\
I    & 1.343	&   1.819   &  55   \\
\hline                  
\end{tabular}
\end{center}
The measurements are from Doroshenko et al. (2005), photometric data, the parameters of the disk/corona model are as in Fig.~\ref{fig:no_wa}, the fourth column gives the amount of starlight required to dilute the variability to match the observations. 
\label{tab:rms}
\end{table}

\section{Discussion}
\label{sect:discussion}

We considered two geometries that can provide the required efficient 
irradiation of the outer parts of an accretion disk responsible for the optical 
emission: a disk/warm absorber geometry and a disk/corona geometry. In the first 
case the scattering material is located relatively far from the disk, around 
the symmetry axis, while in the second geometry the scattering material is 
located closer to the disk surface.

\begin{figure}
   \includegraphics[width=8.5cm]{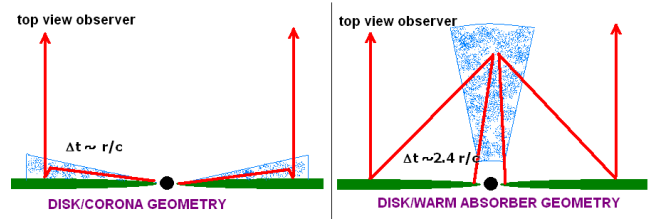}
      \caption{Schematic illustration why the time delays are systematically larger in the disk/warm absorber geometry of the scattering medium than in the disk/corona geometry.
              }
         \label{fig:schematic}
   \end{figure}

We compared the predictions of these two geometries to the measurements of the 
time delays between the various continuum bands. The expected time delays are
generally shorte by a factor of 2-3 in the case of the disk/corona geometry 
than in the disk/warm absorber geometry. This is due to the fact that 
for a disk radius of interest, $r$, the
effectively scattering coronal region is located roughly at the disk radius
distance from the center while the most effectively scattering warm absorber 
region is located at the distance $r$ along the symmetry axis and the overall
distance traveled by photons is roughly $r + \sqrt{r} = 2.4 r$ before the
radiation reaches the disk surface, as illustrated in Fig.~\ref{fig:schematic}.

 The rather short time delays 
measured by Sergeev et al. (2005) strongly favored disk/corona geometry and 
low inclination of the source. High inclination of this source 
($i = 58^{+1}_{-4}$), suggested by Crummy et al. (2006) on the basis of the 
reflection model of the soft X-ray excess, is ruled out at the 2-$\sigma$ 
level. Both models in their simple versions cannot account for a 
decrease of the variability with wavelength since the irradiation, needed to
reproduce the observed spectra redder than a bare disk spectrum, also increases
with the wavelength. However, our simple approach to the wavelength-dependent 
variability amplitude did not take into account the increase of the 
reprocessing region with the wavelength which may result in additional 
suppression of the variability in R and I bands. The possibility of a starlight
contribution alone did not seem to provide the solution.

The disk/corona set of models favored by the delay measurement of Sergeev 
et al. (2005) have black hole masses in the range from $10^7 M_{\odot}$ to
$3 \times 10^7 M_{\odot}$. Such values are in agreement with the best mass
measurement based on reverberation ($M = 1.42 \pm 0.37 \times 10^7 
M_{\odot}$, Peterson et al. 2004). The acceptable accretion rates 
cover a broad range (see Table~\ref{tab:cosi2}). 
However, if we adopt the Peterson et al. value of the mass, the required accretion rate is within the
range of  0.55 - 0.75 (in dimensionless units). 
This is consistent with the typical
values derived for NLS1 objects (e.g. Kuraszkiewicz et al. 2000). The other
parameters of the model, required by fitting the continuum, are also 
reasonable. The parameter $\gamma$ in all acceptable models was 1.8, 
and the fraction $f$ of the bolometric luminosity undergoing reprocessing
should be between 0.12 (for an accretion rate of 0.55) and 0.07 (for an
accretion rate 0.75), for the Peterson et al. mass. On the other hand, if indeed
a small inclination angle is favored for this source, the estimate of the
mass from reverberation may be strongly biased (see the discussion by 
Collin et al. 2006).

The corona model is roughly consistent with the expectations of the Inverse
Compton heated disk/coronae and disk/wind scenarios developed in a number of 
papers (Begelman, McKee \& Shields 1983, Ostriker et al. 1991, 
Murray et al. 1994; see Woods et al.
1996 for hydrodynamical simulations). The inner disk develops a static corona
with negligible thermal outflow while the outer region, where the Inverse 
Compton temperature is larger than the local virial temperature, develops a 
strong wind. The transition region is estimated to be at a fraction of the 
Inverse Compton radius (maximum strength of the corona at 0.2 in the 
simulations of Woods et al.). 
The disk optical/UV spectra obtained in 
the present paper 
are similar to those determined from
more exact coronal solutions (e.g. Kurpiewski et al. 1997). Assuming an 
Inverse Compton temperature corresponding to Mrk 335 of 
$1.3 \times 10^7$ K, we 
obtain the predicted 
outer radius of the corona of $4 \times 10^4 R_{Schw}$, of the same order
as required by our fits to the data. 

The disk/warm absorber geometry is not consistent 
with the delay measurements of Sergeev et al. (2005). 
This indicates that there is no massive outflow of the fully 
ionized material along our line of sight to Mrk 335, although 
such an outflow of considerable optical depth was suggested to take place 
in another NLS1 galaxy (PG1211+143, King \& Pounds 2003). Our results are
 consistent with the values 
obtained for the partially ionized warm absorber column in Mrk 335 
(small value of the column density $N_H =4.2^{+0.7}_{-0.3} \times 10^{22}$ cm$^{-2}$, 
equivalent to $\tau = 0.028$, Bianchi et al. 2001; a similar value of 
$N_H =2.5 \times 10^{22}$ cm$^{-2}$ was obtained by Longinotti et al. 2006),  
since such a warm absorber would not backscatter a significant amount
 of radiation.  Unfortunately, for PG1211+143, no delay measurements exist, 
 so we cannot perform consistency tests for that source.

\begin{acknowledgements}
We thank Arkadiusz Olech and Valya Doroshenko for very helpful discussions,
 Zbyszek Loska for his 3-D computations of the scattering of photons by the 
corona, and Suzy Collin for many helpful remarks which helped to improve the
original version of the manuscript significantly.
Part of this work was supported by grant 
1P03D00829 of the Polish
State Committee for Scientific Research. This research has made use of the NASA/IPAC Extragalactic Database (NED) which is operated by the Jet Propulsion Laboratory, California Institute of Technology, under contract with the National Aeronautics and Space Administration.
\end{acknowledgements}

\end{document}